\documentclass[onecolumn,floatfix,aps,nofootinbib]{revtex4-1}
\usepackage[utf8]{inputenc}
\usepackage{array}
\usepackage{amsmath}
\usepackage{amsfonts}
\usepackage{amssymb} 
\usepackage{wasysym}
\usepackage[all,cmtip]{xy}
\usepackage[english]{babel}
\usepackage{bbm}
\usepackage{caption}
\usepackage{verbatim,upgreek}
\usepackage{array}
\usepackage{amsmath}
\usepackage{multirow}
\usepackage{amsfonts}
\usepackage{amssymb}
\usepackage{slashed} 
\usepackage{hyperref}
\usepackage{leading}
\usepackage[dvipsnames]{xcolor}
\hypersetup{colorlinks=true,linkbordercolor=Blue,linkcolor=Blue, citecolor=Blue}
\usepackage{subcaption}
\usepackage{graphicx}
\usepackage{subeqnarray}
\usepackage{setspace}
\usepackage{indentfirst} 

\usepackage{gensymb}
\newsavebox\myboxA
\newsavebox\myboxB
\newlength\mylenA
\newcommand{\overbar}[1]{\mkern 1.5mu\overline{\mkern-1.5mu#1\mkern-1.5mu}\mkern 1.5mu}
\begin{document}

\title{Exotic fermionic fields and minimal length}

\author{J. M. Hoff da Silva$^{1}$} \email{julio.hoff@unesp.br}
\author{D. Beghetto$^{1,2}$} \email{dbeghetto@feg.unesp.br}
\author{R. T. Cavalcanti$^{1}$}\email{rogerio.cavalcanti@unesp.br}
\author{R. da Rocha$^{3}$}\email{roldao.rocha@ufabc.edu.br} 

\affiliation{$^{1}$Departamento de F\'isica, Universidade
Estadual Paulista, UNESP, Av. Dr. Ariberto Pereira da Cunha, 333, Guaratinguet\'a, SP,
Brazil.}
\affiliation{$^{2}$ Instituto Federal de Educa\c c\~ao, Ci\^encia e Tecnologia do Norte de Minas Gerais, IFNMG,\\ Av. Dr. Humberto Mallard, 1355, Pirapora, MG, Brazil.}
\affiliation{$^{3}$CMCC,  Federal University of ABC, 09210-580, Santo Andr\'e, Brazil.}

\begin{abstract}
We investigate the effective Dirac equation, corrected by merging two scenarios that are expected to emerge towards the quantum gravity scale. Namely, the existence of a minimal length, implemented by the generalized uncertainty principle, and exotic spinors, associated with any non-trivial topology equipping the spacetime manifold. We show that the free fermionic dynamical equations, within the context of a minimal length, just allow for trivial solutions, a feature that is not shared by dynamical equations  for  exotic spinors. In fact, in this coalescing setup, the exoticity is shown to prevent the Dirac operator to be injective, allowing the existence of non-trivial solutions. 
\end{abstract}

\maketitle

\section{Introduction}

Over the last decades, high energy physics has allowed us to look closer to the very structure of matter.  It naturally rises concerns on the limits of current  approaches. Whether this limit exists or not, fundamental concepts regarding the underlying spacetime geometry may be revisited,  at high energy scales \cite{pri,seg}.  At least at energy scales as large as the Planck scale,  quantum gravity effects are expected to set in. The non-renormalizability of quantum gravity consists of a major problem that physicists have been trying to overcome for decades. However,  it could be effectively circumvented, by suggesting that gravity should lead to an ultraviolet cutoff, leading to a minimal observable length.  The existence of a minimal length scale seems to be a model-independent feature of quantum gravity \cite{Kuntz:2019gup}, from string theory \cite{Gross:1987ar,Konishi:1989wk,Amati:1988tn} to loop quantum gravity \cite{Rovelli:1994ge}, and quantum black holes \cite{Casadio:2017sze,Kuntz:2019gka,Scardigli:1999jh,Maggiore:1993rv,Casadio:2015jha,Iorio:2019wtn}. Effectively, the Heisenberg Uncertainty Principle (HUP) can be corrected by a minimal length, giving rise to the so-called {Generalized Uncertainty Principle} (GUP).  See Refs. \cite{Hossenfelder:2012jw,Sprenger:2012uc,Tawfik:2015rva,Casadio2015b} for a comprehensive review. In particular, in the context of the Dirac equation, phenomenological bounds were explored in Ref. \cite{Scardigli:2014qka}. %

Formally, the minimal length can be implemented in quantum mechanics,  through small quadratic corrections to the canonical commutation relations \cite{Kempf:1993bq,Kempf:1994su,Hinrichsen:1995mf,Kempf:1996fz}, whose relationship to a minimal length is not biunivocal \cite{Bishop:2019yft}. The compatibility with relativistic covariance  was implemented in Ref. \cite{QM}, with the so-called Quesne-Tkachuk algebra, leading to a generalization of the Heisenberg algebra. Some high energy physics phenomena were investigated in the scope of the Quesne-Tkachuk algebra \cite{b,tur}, whereas  generalizations were proposed in Ref. \cite{gp}. Besides, the relationship between a  generalized algebra, induced by a minimal length, and higher derivative theories was established \cite{no}. Interacting fermionic fields were also studied in the context of a minimal length scale. Reciprocal effects of massive fermions were investigated in Ref. \cite{osci} and black hole tunnelling was scrutinized   in Ref. \cite{bh}. Interestingly, the investigation of free (classical) fermionic fields whose dynamics is dictated by the deformed Dirac operator leads to a particularly odd result, as shown in the next section. The situation may be framed as follows: denoting by $\mathcal{E}$ the space of  the  Dirac equation solutions engendered by the Dirac operator, $\slashed{\partial}$, it is clear that $\ker(\slashed{\partial})=\mathcal{E}$. In the presence of a minimal length, as we will see, within the usual first-order approximation  and considering massive particles,  the Dirac operator becomes injective, collapsing its kernel into the null spinor. In other words, only the trivial solution is allowed. 

A different perspective is reached by studying exotic spinors, along with the minimal length framework. Given a manifold with non-trivial topology, more than one spinor bundle is allowed. Such bundles belong to different equivalence classes, giving rise to the so-called exotic spin structures, which accommodate exotic spinors \cite{mil}. Inequivalent spin structures are felt through different spin connections \cite{2,SC}, inferring a different dynamics. In other words,  these spinors are annihilated by a modified (exotic) Dirac operator. Moreover, going towards the Planck scale, besides the presence of a minimal length, one also finds support  for the existence of exotic spinors, as non-trivial topologies are also expected to appear in such a regime \cite{dr1,dr2}. Coalescing the two scenarios, we will show that the odd feature about free fermions,  within  dynamical equations corrected by minimal length, is solved in the scope of exotic spinors, being the doubly corrected Dirac operator not injective. For completeness, and anticipating one of the results of Section III.A, we also study particular cases for which the trivial solution is also the only possibility, in the coalescing scenario. 

This paper is organized as follows: the next section is reserved to introduce the problem of free fermionic fields, within the minimal length scope. Sect. III is devoted to studying exotic spinors, as well as the merging between exotic spinors and the minimal length setup. The formal aspects are analyzed and a correction in the topological term is proposed, after what the dynamical scenario is explored. In the final section, we summarize and comment on our relevant results.

\section{Introductory remarks on minimal length and free fermionic systems}

The generalized commutation relation between $X^\mu$ and $P^\mu$, engendered by a minimal length \cite{QM} in the momentum space, reads\footnote{Throughout the paper, natural units are used.} 
\begin{equation}
[X^\mu,P^\nu]=-i[(1-\beta p^\rho p_\rho)\eta^{\mu\nu}-2\beta p^\mu p^\nu],\label{ola}
\end{equation} where the (square of the) minimal length is encoded into the $\beta$ parameter. It is possible to show \cite{QM,no} that a quite suitable point in the parameter space may be chosen so that Eq. (\ref{ola}) holds in the configuration space, for $P^\mu=(1-\beta p^\rho p_\rho)p^\mu$. Hence a correction in the partial derivative operator can be implemented by  
\begin{equation}
\partial_\mu \mapsto (1+\beta \Box)\partial_\mu. \label{ola2}
\end{equation}
With Eq. (\ref{ola2}), the investigation of the free spinors profile,
 in a base manifold endowed with minimal length, follows straightforwardly. The minimal length derivative prescription simply yields 
\begin{equation}
i(1+\beta\Box)\slashed{\partial}\psi -m\psi=0. \label{u1}
\end{equation} Inserting the operator $[i(1+\beta\Box)\slashed{\partial}+m1_{4\times 4}]$ into Eq. (\ref{u1}) from the left,  and keeping the first-order approximation in $\beta$,  one has 
\begin{equation}
\Box\psi+2\beta\Box^2 \psi+m^2\psi=0. \label{u2}
\end{equation} Replacing back $\Box\psi$, coming from (\ref{u2}), into Eq. (\ref{u1}), implies that 
\begin{equation}
i(1-\beta m^2)\slashed{\partial}\psi-m\psi=0. \label{u3}
\end{equation} To make explicit the degenerate states corresponding to the two spin projections, the usual approach suggests the analysis of Eq. (\ref{u3}) in the rest frame formalism, in the momentum space (the Dirac representation of the gamma matrices is particularly clear in this regard), and proceed with a spinorial boost. Writing $\psi=e^{\mp ipx}w(\vec{p})$, where $w(\vec{p})$ is a spinor, it is straightforward to see  that, in the rest frame, Eq. (\ref{u3}) leads to 
\begin{align} 
\left(
\begin{array}{cccc}
 \kappa-1 & 0 & 0 & 0 \\
 0 & \kappa-1 & 0 & 0 \\
 0 & 0 & -\kappa-1 & 0 \\
 0 & 0 & 0 & -\kappa-1 \\
\end{array}
\right)w(\vec{p})=0,\label{u4}
\end{align} for $\kappa\equiv \pm(1-\beta m^2)$. Since $\kappa\neq\pm 1$ for $\beta\neq 0$, the unique solution consists of the trivial one, despite the matrix in Eq. (\ref{u4}) is  not invertible. One could argue that the rest frame would not be attainable, however, then we are forced to conclude that the dynamical equation is insensitive to the minimal length (see, for instance, Eq. (\ref{u3}) for $m=0$). For this reason, henceforth, we will no longer deal with the non-massive case. Following a different approach, and starting with a different corrected Dirac equation, in Ref. \cite{Nozari:2005ix} the authors arrive at a similar conclusion. The existence of a minimal length does not allow the existence of free fermions solutions.

Returning to the analysis performed in the Introduction, let $\psi$ be a section of the $P_{Spin_{(1,3)}}\times \mathbb{C}^4$ spin-Clifford bundle. Hence, denoting by $\mathcal{E} \subset P_{Spin_{(1,3)}}\times \mathbb{C}^4$ the subset comprising solutions of the Dirac equation engendered by the Dirac operator, $\slashed{\partial}$, one sees that in the context of minimal length, the kernel of $\slashed{\partial}$ is given by $\{0\} \subset \mathcal{E}$, and the operator becomes injective. It is difficult to encompass such behavior in a physical interpretation. On the one hand, it seems that the energy increment, to the point where the minimal length becomes dominant, would implement a mapping  $\ker(\slashed{\partial})=\mathcal{E} = \{0\}$. However, as the energy increases, the second-order terms start to be relevant, at least compared to  first-order terms, as these terms contrast to the usual terms. In this case, Eqs. (\ref{u1} -- \ref{u3}) would be modified accordingly, preventing $\ker(\slashed{\partial})$ to collapse into the subset $\{0\}$. The peculiar point is that usual solutions would need second-order terms to survive, whereas non-trivial exotic fermionic solutions, as we will see, do exist in the first-order approximation. To make our claims clear, looking for low energy effects, corrections of the dynamical equations will be restricted to the leading order.  

\section{Exotic spinor structures and the minimal length}

 We start with a discussion on pertinent details, concerning the foundations and construction of exotic spinors, which are particularly important for our analysis. 

A spin structure on a 4-dimensional Lorentzian spacetime manifold $M$ requires an orthogonal frame bundle $
P_{Spin(1,3)} \xrightarrow{\uppi_\mathfrak{s}} M$ and a double cover \begin{equation}
\mathfrak{s}: P_{Spin(1,3)} \rightarrow P_{SO(1,3)},\end{equation} such that $\uppi_\mathfrak{s} = \uppi \circ \mathfrak{s}$, for both $\uppi: P_{SO(1,3)} \rightarrow M$ and $\uppi_\mathfrak{s}$ denoting the bundle projection operator onto the $M$ manifold. When the first cohomology group is not trivial, $H^1(M,\mathbb{Z}_2) \neq \{0\}$,  exotic spin structures $(\tilde{P}_{Spin(1,3)}, \tilde{\mathfrak{s}})$, inequivalent to the usual ones, are then admitted to exist. Hence, one can construct exotic spinors as sections of the spinor bundle, associated to the principal bundle $\tilde{P}_{Spin(1,3)}$. 

In order to evince the difference between the usual and the exotic spin structures, and making explicit the non-trivial elements of $H^1(M,\mathbb{Z}_2)$, let one remembers that two spin structures $P:=(P_{Spin(1,3)}, \mathfrak{s})$ and $\tilde{P}:=(\tilde{P}_{Spin(1,3)}, \tilde{\mathfrak{s}})$ are equivalent as long as a $Spin(1,3)$-equivariant mapping  $q:P \rightarrow \tilde{P}$, compatible to ${\mathfrak{s}}$ and $\tilde{\mathfrak{s}}$, exist. Namely, this equivalence holds if the diagram
	\begin{equation}
	    \xymatrix{
	      P \ar[rdd]_{\mathfrak{s}} \ar[rr]^q & & \tilde{P} \ar[ldd]^{\tilde{\mathfrak{s}}}\nonumber \\
	      & & \\
	      & P_{SO(1,3)}
	    }
	 \end{equation}
	commutes. 
The group homomorphism $\upsigma: Spin(1,3) \rightarrow SO(1,3)$ has $\ker(\upsigma) \simeq \mathbb{Z}_2$. Let $\cup_{i \in I} U_i$ be an open cover for $M$, having transition functions 
	\begin{equation}
	 \mathfrak{a}_{ij}: U_i \cap U_j \rightarrow SO(1,3),
	\end{equation}
	such that $\mathfrak{a}_{ij} \circ \mathfrak{a}_{jk} = \mathfrak{a}_{ik}$ on $U_i \cap U_j \cap U_k$ and $\mathfrak{a}_{jj} = \mathrm{id}$. For a spin structure $(P_{Spin(1,3)}, \mathfrak{s})$ on $M$, a system of transition functions
	\begin{equation}
	 \mathfrak{b}_{ij}: U_i \cap U_j \rightarrow Spin(1,3)
	\end{equation}
	has the  properties \cite{petry,2}
	\begin{equation}
	 \upsigma \circ \mathfrak{b}_{ij} = \mathfrak{a}_{ij},\qquad\mathfrak{b}_{ij} \circ \mathfrak{b}_{jk} = \mathfrak{b}_{ik},\qquad \mathfrak{b}_{jj} = \mathrm{id}.
	\end{equation}
	In this way, two spin structures $(P_{Spin(1,3)}, \mathfrak{s})$ and $(\tilde{P}_{Spin(1,3)}, \tilde{\mathfrak{s}})$ are, respectively, portrayed by the mappings $\mathfrak{b}_{ij}$ and $\tilde{\mathfrak{b}}_{ij}$, being both defined on $U_i \cap U_j$ to $Spin(1,3) = \widetilde{Spin}(1,3)$, such that $\upsigma \circ \mathfrak{b}_{jk} = \mathfrak{a}_{jk} = \upsigma \circ \tilde{b}_{jk}$.
	
	Let one defines a cocycle  $\mathfrak{c}_{ij}$  by the expression $\mathfrak{b}_{ij}(x) = \tilde{\mathfrak{b}}_{ij}(x) \mathfrak{c}_{ij}$, such that
	\begin{equation}
	 \mathfrak{c}_{ij}: U_i \cap U_j \rightarrow \ker(\upsigma)=\mathbb{Z}_2 \hookrightarrow Spin(1,3),
	\end{equation}
	with $\mathfrak{c}_{ij} \circ \mathfrak{c}_{jk} = \mathfrak{c}_{ik}$. These cocycles $\mathfrak{c}_{ij}$ are the non-trivial elements of the first cohomology group $H^1(M,\mathbb{Z}_2)$. This construction is shown to define an one-to-one correspondence, between inequivalent spin structures and $H^1(M,\mathbb{Z}_2)$ \cite{3}. This shows that the very existence of the cocycles $\mathfrak{c}_{ij}$ is necessary to have exotic spinors on $M$. Yet, notice that $\mathfrak{c}_{ij}$ requires the existence of an open covering for $M$.

	Now, the exotic spinors construction on a manifold will be addressed, $\widetilde{\overbar{M}}$, endowed with both non-trivial topology and minimal length. We have already seen that to establish exotic spinors on a manifold, one needs an open cover  $\cup_{i \in I} U_i$ for $\widetilde{\overbar{M}}$. On the other hand, in the current case it is necessary to proceed with some caution, due to the minimal length. More precisely, the bound \begin{equation}|U_i \cap U_j| > L\end{equation} is necessary, for all $i,j$, where $L$ accounts for the minimal length. Since $\widetilde{\overbar{M}}$ is a Hausdorff paracompact topological space, then every open cover has an open refinement that is locally finite, which ensures the possibility of such open sets to exist. In other words, as the arbitrary union of open sets is also an open set,  it is always possible to construct a collection $W$ of open sets,  $W_k$, implemented by an arbitrary union of $U_i$ as   $W_k=\cup_i U_i$, for all $k$, such that $W=\cup_k W_k$. The  $W_k$ are chosen such that $|W_k|>L$, for all $k$, and $|W_k\cap W_l|>L$, for all $k,l$, covering the manifold. Hence, the construction of the transition functions $\overbar{\mathfrak{a}}_{ij}: W_i \cap W_j \rightarrow \overbar{SO}(1,3)$ is allowed. Therefore, also $\overbar{\mathfrak{h}}_{ij}: W_i \cap W_j \rightarrow \overbar{Spin}(1,3)$ and $\tilde{\overbar{\mathfrak{h}}}_{ij}: W_i \cap W_j \rightarrow \widetilde{\overbar{Spin}}(1,3)$ can be deployed, such that $\upsigma \circ \overbar{\mathfrak{h}}_{ij} = \overbar{\mathfrak{a}}_{ij},$ $\overbar{\mathfrak{h}}_{ij} \circ \overbar{\mathfrak{h}}_{jk} = \overbar{\mathfrak{h}}_{ik}$, and  $\overbar{\mathfrak{h}}_{jj} = \mathrm{id}$. Here, $\upsigma$ is a group homomorphism between $Spin(1,3)=\overbar{Spin}(1,3) = \widetilde{\overbar{Spin}}(1,3)$ and $\overbar{SO}(1,3)$.
	
	As a conclusion of this construction, it is always possible to find cocycles $\overbar{\mathfrak{c}}_{ij}: W_i \cap W_j \rightarrow \mathbb{Z}_2$, defined by $\overbar{\mathfrak{h}}_{ij}(x) = \tilde{\overbar{\mathfrak{h}}}_{ij}(x) \overbar{c}_{ij}$, which are non-trivial elements of $H^1\left(\widetilde{\overbar{M}},\mathbb{Z}_2\right)$, allowing the existence of exotic spinors on $\widetilde{\overbar{M}}$. 

We will study, in the next section, the behavior of exotic spinors, $\tilde{\psi}$, along with the minimal length corrections in their dynamics. It is worth to emphasize that the correction due to the non-trivial topology is implemented when one  replaces the Dirac operator as 
\begin{eqnarray}
\slashed{\partial}\mapsto \slashed{\partial} + \upchi^{-1} d\upchi. \label{la}
\end{eqnarray} Therefore this topological correction encompasses  a derivative term which could, in principle, also be corrected by the minimal length. Let us look at this issue more closely. %
Following Refs. \cite{SC,avis0,isham0,isham1,petry}, one assumes a set of scalar fields $\upchi_i: U_i\rightarrow\mathbb{C}$
  such that $\upchi_i(x)\in $ U(1), and
\begin{align}\label{xi1}
 \upchi_i(x) (\upchi_j(x))^{-1} = \omega(\mathfrak{c}_{ij}(x)) = \pm 1, 
  \end{align}  
  where $\omega$ denotes a faithful irreducible representation from the Minkowski spacetime Clifford algebra to the space $M(4,\mathbb{C})$
  of $4\times 4$ matrices with complex entries. 
If the case of a 2-torsionless second cohomology group $H^2(M,\mathbb{Z}_2)$, the $\upchi_i(x)$ scalar fields always exist, and ${\upchi}^2_i(x) = {\upchi}^2_j(x)$, for $x\in U_i\cap U_j$  \cite{isham0,isham1,petry,thomas}. Consequently, the local functions $\upchi_i$ define a unique unimodular scalar field $\upchi:M\rightarrow\mathbb{C}$ such that, for all $x\in U_i$, the equality $\upchi(x) = {\upchi}_i^2(x)$ holds.

Besides, the $\upchi_i$ scalar field forms a local root system for $\upchi$ \cite{petry}, and are the generators of the cocycles $\overbar{c}_{ij}$, which are the non-trivial elements of $H^1(M,\mathbb{Z}_2)$. As already presented, such cocycles do exist and,  equivalently, exotic spinors exist in this new spacetime manifold, so it seems natural to think that their generators carry information on the minimal length that characterizes $\widetilde{\overbar{M}}$. Yet, the unimodular scalar fields may be arranged  as 
\begin{equation}
\upchi(x)=e^{i\uptheta(x)},
\end{equation} where $\uptheta: M \rightarrow \mathbb{R}$ \cite{petry}. Usually, without a minimal length, the $\uptheta(x)$ represents a strongly continuous function that reflects the non-trivial topology. Within the minimal length context, however, $|\upchi(x')-\upchi(x)|=2-2\cos(\uptheta(x')-\uptheta(x))$, and one is certainly forbidden to make $x'$ infinitely close of $x$. Equivalently, $|\upchi(x')-\upchi(x)|\sim O(\beta)$, where $\beta$ has $L^2$ order of magnitude, and we are forced to conclude that the derivative operator should also be corrected. Nevertheless, the formal and exact functional form of such a correction is a quite difficult task. Up to the best of our knowledge, there is no completely satisfactory answer to this issue, culminating in a manageable operator. Here we will adopt, instead, an effective approach, motivating the corrected derivative operator. Our proposal goes as follows: since the minimal length is  responsible for this correction in the derivative operator, it must be small compared to the identity, 
\begin{equation}
\partial\mapsto (1_{4\times 4}+\tilde{\beta}\mathcal{O})\partial,
\end{equation} 
 where hence $\tilde{\beta}\sim \beta$, and $\mathcal{O}$ is an operator acting on sections of $\widetilde{\overbar{P_{Spin(1,3)}}}\times\mathbb{C}^4$. Now, nothing ensures that $\mathcal{O}$ is a differential operator. If it is not, one would have to take special care ensuring covariance of the dynamical equation. Being $\mathcal{O}$ a derivative operator, however, covariance imposes that either $\mathcal{O}=a_0\gamma^\mu\partial_\mu$ or it contains only even powers in the derivative. In this last case, its more complete form would be given by $\mathcal{O}=a_1\Box+a_2\Box^2+\cdots$, where $a_0, a_1, a_2,\cdots$ are arbitrary parameters with suitable dimensions. Anticipating Sec. III.A, in the first order approximation there is no substantial difference in the net result. We will present our calculations with $\mathcal{O}$ being taken as a derivative operator of the form $\mathcal{O}=\Box$, with $a_1$ already absorbed into $\tilde{\beta}$, and just mention the other case by passing. This motivated functional form of correction is, then, quite similar to  the minimal length correction in the derivative operator obtained, however, by other arguments. 

As a parenthetical remark, we notice that the topological term is commonly accepted to be eliminated by a gauge redefinition. We will end this section making explicit that this cannot be accomplished \cite{aaca,daSilva:2016htz} (For this remark we will not pay attention to minimal length issues).  The explicit form of Eq. (\ref{la}), along with a gauge interaction term (obtained for instance by the minimal coupling prescription), is given by \begin{equation} (i\gamma^\mu\partial_\mu+\gamma^\mu A_\mu+i\gamma^\mu\partial_\mu\uptheta(x)-m1_{4\times4})\psi=0.
\end{equation}  As a connected group, the $U(1)$ representations are unitary and may be disposed as $e^{i\Uplambda(x)}$, with real $\Uplambda(x)$. By means of a gauge transformation $A_\mu\mapsto A_\mu-\partial_\mu\Uplambda(x)$ it would be necessary the identification of $\Uplambda(x)$ with $i\uptheta(x)$, to eliminate the topological term. Nevertheless, $\uptheta(x)$ is also real and, therefore, such elimination cannot be accomplished.

\subsection{Exotic fermions and minimal length}

Bearing in mind the discussion performed in the previous section, we will  investigate the free fermionic exotic case, with a minimal length correction in the usual derivative and also with the correction in the exotic term. This last correction is not so determinant after all, and its absence would not significantly modify the final result. At the end of the analysis, we also comment about non-derivative corrections. 

The exotic spinor dynamical equation, in the context here explored reads
\begin{eqnarray}
i\gamma^\mu\left(1+\beta \Box\right)\partial_\mu\tilde{\psi}+i\gamma^\mu\left(1+\tilde{\beta}\Box\right)\partial_\mu\uptheta(x)\tilde{\psi}-m\tilde{\psi}=0.\label{exo1}
\end{eqnarray} It is helpful to remember that the corrected derivative terms are already saturated, that is $\Box\partial_\mu\cdot=\partial^\rho[\partial_\rho(\partial_\mu\cdot)]$. For further use, we write Eq. (\ref{exo1}) in the form
\begin{eqnarray}
i\slashed{\partial}\tilde{\psi}+i\slashed{\partial}\uptheta(x)\tilde{\psi}+i\slashed{\partial}(\beta\Box\tilde{\psi})+i\tilde{\beta}\Box\slashed{\partial}\uptheta(x)\tilde{\psi}-m\tilde{\psi}=0. \label{exo2} 
\end{eqnarray} Inserting from the left the operator $\left[i(1+\beta \Box)\slashed{\partial}+i\left(1+\tilde{\beta}\Box\right)\slashed{\partial}\uptheta(x)+m1_{4\times 4}\right]$ into Eq. (\ref{exo2}),  we arrive at the following awkward expression, 
\begin{eqnarray}
-\Box\tilde{\psi}-2\beta\Box^2\tilde{\psi}-m^2\tilde{\psi}&+&\left.i\left[i\slashed{\partial}+\beta\Box\slashed{\partial}+m1_{4\times 4}\right]\left[\slashed{\partial}\uptheta(x)\tilde{\psi}+\tilde{\beta}\Box\slashed{\partial}\uptheta(x)\tilde{\psi}\right]+
i\left[\slashed{\partial}\uptheta(x)-\tilde{\beta}\Box\slashed{\partial}\uptheta(x)\right]\left[i\slashed{\partial}\tilde{\psi}+i\beta\Box\slashed{\partial}\tilde{\psi}-m\tilde{\psi}\right]\right.\nonumber\\&-&\left. \left[\slashed{\partial}\uptheta(x)+\tilde{\beta}\Box\slashed{\partial}\uptheta(x)\right]\left[\slashed{\partial}\uptheta(x)\tilde{\psi}+\tilde{\beta}\Box\slashed{\partial}\uptheta(x)\tilde{\psi}\right]=0.\right. \label{exo3} 
\end{eqnarray} Notice that, in Eq. (\ref{exo2}), $\Box\tilde{\psi}$ is already multiplied by $\beta$. Therefore, in the first-order limit, and considering $\beta\tilde{\beta}\rightarrow 0$, as discussed, the relevant terms (in advance of what will  be used back into Eq. (\ref{exo2})) read\footnote{In this regard, it is possible to see that the only term encoding the minimal length correction in the exotic term is already present in Eq. (\ref{exo2}), in the first-order approximation.}
\begin{eqnarray}
-\beta\Box\tilde{\psi}-\beta m^2\tilde{\psi}-\beta\slashed{\partial}(\slashed{\partial}\uptheta(x)\tilde{\psi})-\beta\slashed{\partial}\uptheta(x)\slashed{\partial}\tilde{\psi}-
\beta\slashed{\partial}\uptheta(x)\slashed{\partial}\uptheta(x)\tilde{\psi}=0.\label{exo4}
\end{eqnarray} The only tricky term is $\slashed{\partial}(\slashed{\partial}\uptheta(x)\tilde{\psi})=
\Box\uptheta(x)\tilde{\psi}+\gamma^\mu\gamma^\nu\partial_\nu\uptheta(x)\partial_\mu\tilde{\psi}$. Using $\{\gamma^\mu,\gamma^\nu\}=2\eta^{\mu\nu}$ implies that 
\begin{equation}
-\beta\slashed{\partial}(\slashed{\partial}\uptheta(x)\tilde{\psi})=-\beta\Box\uptheta(x)\tilde{\psi}-2\beta\partial^\mu\uptheta(x)
\partial_\mu\tilde{\psi}+\beta\slashed{\partial}\uptheta(x)\slashed{\partial}\tilde{\psi}. \label{exo5}
\end{equation} Inserting back this expression into Eq. (\ref{exo4}) yields  
\begin{eqnarray}
\beta\Box\tilde{\psi}=-\beta m^2\tilde{\psi}-\beta\Box\uptheta(x)\tilde{\psi}-2\beta\partial^\mu\uptheta(x)\partial_\mu\tilde{\psi}-
\beta\partial^\mu\uptheta(x)\partial_{\mu}\uptheta(x)\tilde{\psi}. \label{exo6}
\end{eqnarray}

Now one can return to Eq. (\ref{exo2}) inserting there $\beta\Box\tilde{\psi}$ obtained above. A bit of simple, but tedious, algebra leads to 
\begin{eqnarray}
i\Big[(1-\beta m^2)-\beta\Box\uptheta(x)\!\!&-&\left.\!\!\beta\partial^\mu\uptheta(x)\partial_\mu\uptheta(x)-2\beta\partial^\mu\uptheta(x)
\partial_\mu\Big]\slashed{\partial}\tilde{\psi}\right.\nonumber\\&+&\left.\Big[i\slashed{\partial}\uptheta(x)+i(\tilde{\beta}-\beta)\Box
\slashed{\partial}\uptheta(x)-2i\beta\slashed{\partial}(\partial^\mu\uptheta(x))\partial_\mu-m1_{4\times4}\Big]\tilde{\psi}=0.\right. \label{exo7}
\end{eqnarray} Before going any further, we remark that Eq. (\ref{exo7})  recovers, in suitable limits, the particular cases it generalizes: 
\begin{itemize}
\item[a)] for $\uptheta(x)$ constant and $\beta=0=\tilde{\beta}$, Eq. (\ref{exo7}) recovers the standard Dirac equation; 
\item[b)] for $\uptheta(x)$ constant and $\beta,\tilde{\beta}\neq 0$, Eq. (\ref{u3}) is obtained;
\item[c)] keeping the function $\uptheta(x)$ and setting $\beta=0=\tilde{\beta}$, the dynamical equation for exotic spinors is attained, as expected. 
\end{itemize} 
In the approximation considered, it is clear that Eq. (\ref{exo7}) circumvents the problem pointed out for Eq. (\ref{u1}). In fact, recasting Eq. (\ref{exo7}) as $\tilde{\slashed{\partial}}\tilde{\psi}=0$, and denoting by $\tilde{\mathcal{E}}\subset\widetilde{\overbar{P_{Spin(1,3)}}}\times\mathbb{C}^4$ the set comprising the solutions of $\tilde{\slashed{\partial}}\tilde{\psi}=0$, then obviously $\ker(\tilde{\slashed{\partial}})=\tilde{\mathcal{E}}\neq \{0\}$. Hence $\tilde{\slashed{\partial}}$ is not injective. The exoticity prevents the minimal length correction to shrinking the kernel of $\tilde{\slashed{\partial}}$. Non-trivial exotic solutions for the non-interacting problem do survive, in fact. 

In view of the results  pointed here and the problem previously described, one would be tempted to think of the minimal length setup as being the natural ambiance to find out physical signatures exclusively implemented to exotic spinors. Nevertheless, keeping in mind that our investigation was carried out focusing on free spinors, such assertion is not as straightforward. What seems plausible to conclude is that physical signatures regarding exotic spinors will  be dominant (though not exclusive) in a limit of the interacting terms going to zero. However, this phenomenological bridge needs a deeper investigation. In any case, the spectrum arising from Eq. (\ref{exo7}) will  furnish the clues for a typical signature. 

We have asserted that Eq. (\ref{exo7}) admits non-trivial solutions. While it is generally true, we would like to present here a counterexample,  making explicit the fact that for some particular cases of (\ref{exo7}), non-trivial solutions are not allowed. This is important, since it points to the necessity of some caution in drawing conclusions from this fermionic system. In order to make our point clear here, we notice that it is somewhat usual to think of the additional terms in Eq. (\ref{exo7}) as effective interaction terms. Hence, no plane waves are expected. Here this point of view will not be adopted. Instead, the fermionic fields investigated here are completely free, but in a rather unusual (although physically justified) base manifold. If one is willing to accept this approach, then it is plausible to look for plane wave solutions, with possibly a changed energy spectrum. Hence by writing, as before, $\tilde{\psi}=e^{\mp ipx}\tilde{w}(\vec{p})$, Eq. (\ref{exo7}) in the rest frame yields 
\begin{eqnarray}
\pm\Big[1-\beta m^2-\beta\Box\uptheta(x)\!\!&-&\left.\!\!\beta\partial^\mu\uptheta(x)\partial_\mu\uptheta(x)\Big]\gamma^0m \tilde{w}(0)\pm 2i\beta \dot{\uptheta}(x)\gamma^0 m^2 \tilde{w}(0)\right.\nonumber\\&+&\left. \Big[i\slashed{\partial}\uptheta(x)+i(\beta-\tilde{\beta})\Box\slashed{\partial}\uptheta(x)-m\Big]1_{4\times4}\tilde{w}(0)\mp 2\beta \slashed{\partial}(\dot{\uptheta}(x))m\tilde{w}(0)=0.\right. \label{exo8}
\end{eqnarray} In order to make explicit our claim that not every (topologically non-trivial) scenario leads to non-trivial solutions, we investigate the particular case $\uptheta=\uptheta(t)$, i. e., the topological term is only a smooth function of time. In this particular case, Eq. (\ref{exo8}) may be recast as (from now on $\uptheta(t)\equiv \uptheta$)
\begin{eqnarray}
\Big\{\pm \left(1-\beta m-3\beta\ddot{\uptheta}\right)-\beta\dot{\uptheta}^2+i\left[\left(\pm 2m^2\beta+1\right)\dot{\uptheta}+(\tilde{\beta}-\beta)\dddot{\uptheta}\right]\Big\}\gamma^0\tilde{w}(0)-m\tilde{w}(0)=0,\label{exo9}
\end{eqnarray} where in the last term the multiplication by the identity is implicit. Defining $f^{\pm}(\uptheta)=\pm m(1-\beta m-3\beta\ddot{\uptheta}+\beta\dot{\uptheta}^2)$ and $g^{\pm}(\uptheta)=(\pm 2m^2\beta+1)\dot{\uptheta}+(\tilde{\beta}-\beta)\dddot{\uptheta}$,
we may recast Eq. (\ref{exo9}) in $2\times 2$ blocks as 
\begin{align} 
\left(
\begin{array}{cc}
 \big[(f^{\pm}(\uptheta)-m)+ig^{\pm}(\uptheta)\big]1_{2\times 2} & 0_{2\times 2}  \\
 0_{2\times 2}  & \big[(f^{\pm}(\uptheta)+m)+ig^{\pm}(\uptheta)\big]1_{2\times 2}  \\
\end{array}
\right)\tilde{w}(0)=0.\label{exo10}
\end{align} 
Notice that in the $\uptheta=$ constant limit, Eq. \eqref{exo10} recovers (\ref{u4}), as expected. Moreover, besides assuming $\uptheta=$ constant, if $\beta$ and $\tilde{\beta}$ both vanish, then Eq. (\ref{exo10}) turns into the usual Dirac case, recovering the two usual degenerate states for the free particle case. 

As $\uptheta(t)$ is a real function, reflecting the unusual topology of the spacetime, it is clear that $g^{\pm}(\uptheta)$ must be zero, as a necessary (although not sufficient) condition to non-trivial solutions. It is straightforward to see that $g^{\pm}(\uptheta)=0$ leads to $(\tilde{\beta}-\beta)\ddot{\uptheta}+(\pm 2m^2\beta+1)\uptheta=c$, where $c$ is constant\footnote{Note that the case of $\tilde{\beta}=\beta$ yields $\uptheta=$ constant and the premise for topological corrections is undermined.}. The $\uptheta_*$ function, that solves $g^\pm(\uptheta_*)=0$, is then given by a simple combination of sines and cosines. Nevertheless, by inspecting the terms $f^{\pm}(\uptheta)\pm m$, it is readily seen that to have a situation as in the usual Dirac case, it is also necessary to have\footnote{These two conditions together are necessary and sufficient.} $3\ddot{\uptheta}-\dot{\uptheta}^2+m=0$. The solution of this last equation, however, is given by a Weierstrass elliptic function, with no superposition with $\uptheta_*$ for all frequencies. Hence, it is not possible to satisfy all the necessary conditions to have a non-trivial solution. This ends the counterexample. 

We finish by stressing that had we investigated the minimal length correction in the topological term as an operator proportional to $\gamma^\mu\partial_\mu$, let us say $\left(1_{4\times4}+\tilde{\beta}\slashed{\partial}\right)\partial_{\mu}$, the only difference (keeping the same approximation in vogue) would be to replace the term $\tilde{\beta}\Box\slashed{\partial}\uptheta(x)$ by $\tilde{\beta}\Box\uptheta(x)$, in Eq. (\ref{exo7}).

\section{Conclusions}

We have shown that the very existence of a minimal length makes the Dirac operator $\slashed{\partial}$ to be injective, turning its kernel into $\{0\}$, which is not interesting at all for fermionic physics. One can understand, thus, the existence of the minimal length encoded in $\beta$ as a kind of realization of an operation that squeezes the space of solutions of the Dirac equation, $\ker(\slashed{\partial}) = \mathcal{E} \subset P_{Spin_{(1,3)}}\times \mathbb{C}^4$, into a space having only the null spinor. It could be a clue of a feature presented in approaches to quantum gravity. However, when the spacetime admits the existence of exotic spinors, the Dirac equation may have non-trivial spinors as solutions. Thus, assuming that both effects are present in high energy phenomena, the existence of free fermions should depend on which one is effectively dominant when the energy is rising from TeV order towards the Planck scale. It is important to notice, though, that not every scenario merging exotic spinors and a  minimal length allows non-trivial solutions of the Dirac equation. As we have shown in the last section, when one takes the particular case of $\uptheta = \uptheta(t)$, for instance, the kernel of $\tilde{\slashed{\partial}}$ shrinks  to the null spinor again. In other words, it is necessary to have a local exotic term $\uptheta$ (with minimal length), to have non-null spinors. 

The previous discussions indicate an unexpected relationship between the topologies of the spacetime and the space of solutions of the Dirac equation when contributions of a minimal length scale are relevant. This helps to understand which are the feasible physical scenarios for non-null fermionic fields to existing in a spacetime with minimal length. 
Besides, not only the Dirac equation in other contexts \cite{Cianci:2016pvd,Fabbri:2014wda}, but other  first order equations of motion can be explored, driving  flagpole and flag-dipole spinor fields \cite{daRocha:2005ti,daSilva:2012wp,Fabbri:2019kfr}, in the exotic context \cite{Dantas:2015mfi}.  
Further studies on the relationship between these spaces are being carried on by the authors, aiming to expand our understanding of exotic spinors dynamics.

\section{Acknowledgments}
JMHS thanks to CNPq grant No. 303561/2018-1 for partial support. RCT thanks the UNESP-Guaratinguet\'a Graduate program and CAPES for the financial support. RdR is grateful to FAPESP (Grant No. 2017/18897-8) and to the National Council for Scientific and Technological Development -- CNPq (Grants No. 303390/2019-0, No. 406134/2018-9), for partial financial support.

\end{document}